\documentclass[aps,prl.reprint,twocolumn]{revtex4}
\usepackage{epsfig}
\usepackage{graphicx}
\usepackage{amsmath}
\usepackage{amsfonts}
\usepackage{amssymb}
\usepackage{ulem}
\usepackage{cancel}
\usepackage{graphicx}
 \usepackage{wrapfig}
\usepackage[cp1251]{inputenc}
\usepackage{epsfig}
\usepackage{graphicx}
\usepackage{amsmath}
\usepackage{amsfonts}
\usepackage{amssymb}
\usepackage{graphicx}
 \usepackage{wrapfig}
\usepackage{textcomp}
\usepackage{pgf,pgfarrows,pgfnodes,pgfautomata,pgfheaps,pgfshade}
\usepackage{amsmath,amssymb}
\usepackage[colorlinks=true, linkcolor=blue]{hyperref}
\newcommand{{\sign}}{\rm sign}

\begin{document}

\title{
Many-body synchronization of interacting qubits by engineered ac-driving
}

\author{Sergey V. Remizov$^{1,2}$} 
\author{Dmitriy S. Shapiro$^{1,2,3}$}\email{shapiro.dima@gmail.com} \author{Alexey N. Rubtsov$^{1,4,5}$}

\affiliation{$^1$Dukhov Research Institute of Automatics (VNIIA),  Moscow 127055, Russia}
\affiliation{$^2$V. A. Kotel'nikov Institute of Radio Engineering and Electronics, Russian Academy of Sciences, Moscow 125009, Russia}
\affiliation{$^3$National University of Science and Technology MISIS, Leninsky prosp. 4, Moscow, 119049, Russia}
\affiliation{$^4$Russian Quantum Center,   Skolkovo, 143025 Moscow Region, Russia}
\affiliation{$^5$Department of Physics, Moscow State University, 119991 Moscow, Russia}

\begin{abstract}
In this work we introduce the many-body synchronization of an interacting qubit ensemble which   allows one to switch dynamically from many-body-localized (MBL) to an ergodic state.  We show that applying of $\pi$-pulses with altering phases, one can effectively suppress the MBL phase and, hence, eliminate qubits disorder. The findings are based on the analysis of the Loschmidt echo dynamics which shows a transition from a power-law decay to more rapid one indicating the dynamical MBL-to-ergodic transition. The technique does not require  to know the microscopic details of the disorder.
\end{abstract}
\maketitle
\section{Introduction}
Contemporary view of generic properties of quantum ensembles where the interactions and disorder are simultaneously present is that those systems reveal the two states:  the   ergodic state and many-body localized (MBL) phase. The general statement is that  the ergodic, or delocalized,  phase of a closed many-body quantum system  is characterized by  the eigenstate thermalization hypothesis \cite{Srednicki}. It says that the wave function evolution of excitations from a narrow energy domain  is nothing but a relaxation  of observables  to the  Gibbs  distribution  \cite{Deutsch}.
  In contrast, the
MBL phase reveals the non-ergodicity with an absence of an eigenstate thermalization \cite{Pal2010}. It is described by a class of  Hamiltonians where the eigenvalues do not repel and  have Poisson  statistics.
The notation of MBL  was initially introduced in the context of finite temperature metal-insulator transition in a system of  interacting electrons with a static random  potential \cite{Gornyi2005, Basko2006, Oganesyan2007}.  Further, the MBL was  intensively explored  for disordered  spin-1/2 systems, see  \cite{Nandkishore2015} for a review.
One of the remarkable features of MBL phase is that  von Neumann entropy of entanglement  \cite{Bennett1996, Vidal2003} 
  does not show a volume-law scaling for finite temperatures. Instead of that there is an area-law  behavior of the entropy  \cite{Berkovits2012, Bauer2013,BURMISTROV2017140}  which is similar to the low temperature results obtained for the ground state of gapped spin liquids \cite{Vidal2003}.
 The recent experimental observations of MBL phase have been reported for fermions in an optical lattice \cite{Schreiber842}, driven dipolar spin impurities in diamond \cite{Choi2017}  and in a gas of ultracold fermionic potassium atoms \cite{Bordia2017}. Probing of a many-body dynamics on a quantum simulator with controllable 51 $^{87}$Rb atoms  was reported in \cite{Bernien2017}.  Most of the theoretical and numerical realizations of the MBL phase deal with arrays of qubits \cite{Znidaric2008, Bardarson2012,Huse2014,Pal2010,Vasseur2015, Serbyn2013, Serbyn2015}. Their  theoretical models   are of  Heisenberg type  with exchange interaction \cite{Latorre2009, Vitagliano2010} and random qubit frequencies.

 Collective properties of the MBL phase can be probed by a temporal dynamics of the quantum information \cite{Bardarson2012, Serbyn2013, Serbyn2015} stored in a system during its evolution from a particular initial state.
 Namely, the break of ergodicity with the transition into MBL is seen by  a change of the entropy time  dependence  from the linear to logarithmic one. Another   technique   is the Loschmid echo  \cite{Goussev:2012}, being  a quantum mechanical counterpart of the Lyapunov exponent in classical chaotic systems.
  The Loschmidt echo is defined through an overlap of   wave functions which were identical for $t=0$ and start to evolve with two slightly different Hamiltonians. If the system is in  MBL phase then the Loschmidt  echo has a power-law decay, see    \cite{Serbyn2017} and references therein for details. The regime of non-interacting Anderson insulating phase is expected to show a non-vanishing echo while the ergodicity is manifested by a fast exponential decay.

 The external periodic driving  and parametric pumping \cite{Beaudoin2012, Srinivasan2011,Hoffman2011,Chen2014,Zeytinouglu2015,Braumueller2015, Remizov2017,Shapiro2015}
  allow to introduce an additional control over the properties of the superconducting metamaterials realized as qubit ensembles \cite{Macha2014,Shulga2017,fistul2017quantum} and Josephson chains \cite{Zhang2017}. 
 For non-interacting qubits a suppression of an inhomogeneous broadening was demonstrated for the nitrogen-vacancies in diamonds \cite{Dutt2007,Sandner2012,Putz2014}. We later proposed an optimized shape of the driving pulses for the broadening suppression  and studied the effect of strong driving on an entanglement in the hybrid qubit-cavity system \cite{Remizov2015, RemizovShapiroRubtsov2017}. In the context of disordered interacting qubits it was shown that an external driving can enhance the localization effects, namely, a driven transition to the dynamical MBL phase was discovered for an engineered Floquet Hamiltonian  \cite{Lukin}. The main idea of \cite{Lukin} is to impose a phase rotation on each even site of the qubit chain. It allows to effectively suppress the coupling between qubits while preserving the disorder in the qubit frequencies. However this technique   does not seem to allow for an inverse switching from the MBL to an ergodic state, because phase rotations cannot lead to an effective increase of the qubit-qubit coupling.
 
 In this Letter we introduce the many-body synchronization technique allowing for a switching from the MBL to an ergodic state by an effective suppression of the disorder. We demonstrate that the sequence of $\pi$-pulses of an alternating phase, similar to that used in \cite{Putz2014}, leads to the synchronization of the inhomogeneously broadened qubits even in the presence of qubit-qubit interaction. Importantly,  this  technique requires that all qubits are subjected to the same engineered driving, irrespective of a particular disorder realization.

  The paper is organized  as follows. In the Section \ref{model} we present the model of the driven qubits and their non-stationary Hamiltonian. In Section \ref{results} we show the results for a transition from the MBL to ergodic phase   and in \ref{discussion} we conclude.

 \section{Model} \label{model}
   The quantum gates  are the pulse techniques which allow to rotate a vector of the qubit state on the Bloch  sphere on a particular angles. Superconducting technologies offer quantum gates via the coupling between a Josephson qubit  and  GHz transmission line \cite{Yamamoto2010,nation2012colloquium, Blais2004}. 
   The qubit state is  read out by a measuring of a dispersive shift  in an auxiliary  resonator  \cite{Boissonneault2010, George2017}.
   In Ref. \cite{Remizov2015} a  generalized non-rectangular envelope of $\pi$-pulse was proposed. This $\pi$-pulse allows to compensate the effects of disorder in   frequencies in a non-interacting qubit system.  Involving higher number of  harmonics   into the scheme   one can increase a fidelity in excitations of detuned qubits from ground to excited states. In other words, this method allows to  achieve a preassigned accuracy in rotation of qubits' Bloch vectors from south to north pole of the sphere. In this Letter the nearest neighbor Heisenberg exchange interaction is taken into account, while the sequence of $\pi$-pulses  has the rectangular shape. This is nothing but resonant driving with a constant magnitude, which, indeed, has $T$-periodic and an abrupt  altering of its sign.

    The non-stationary  Hamiltonian under consideration is the  following
   \begin{eqnarray}
 	H(t)=\frac{1}{2}\sum\limits_{j=1}^L h_j \sigma^z_j +J\sum\limits_{j=1}^{L-1}  \vec\sigma_j\vec\sigma_{j+1} +\label{hamiltonian-0} \\    +\sum\limits_{j=1}^L\left[ F(t)\sigma^+_j+F^*(t)\sigma^-_j\right] \label{hamiltonian-driving}
   \end{eqnarray}
(Plank constant    $\hbar=1$ throughout the paper).
 Qubits  have a disorder in excitation frequencies which  belong to the following domain $h_j \in [h-W ;h+W]$ and have flat distribution. The unperturbed stationary part of the  Hamiltonian (\ref{hamiltonian-0}) is XXX Heisenberg model with a diagonal disorder. The system described by (\ref{hamiltonian-0}) is known to supports the MBL regime \cite{Oganesyan2007, Pal2010} for $W>W^*$ with $W^*$ is a critical value of the broadening.

    The transversal perturbation term  (\ref{hamiltonian-driving}) has the form of  Zeeman field applied parallel to $xy$-plane in context of the spin models.
    The function of driving $F(t)$ in our consideration is a harmonic signal with  rectangular modulation of an amplitude $f(t)$ which is applied equally to all of the qubits. The carrying frequency of the driving is tuned into the resonance with the average frequency $h$ of the ensemble and $F(t)$ reads as
   \begin{equation}
    F(t)=f(t)\exp({\rm i} h t ).
    \end{equation}
    Periodic  function $f(t)$ is  real  and is  defined for $0<t<T$ as
     \begin{equation}
     	f(t)=
     	\begin{cases}
     		 f, & \text{if }   0<t<T/2, \\
     		-f, & \text{if }   T/2<t<T.
     	\end{cases} \label{f}
     \end{equation}
 The pulse duration time $T/2$ between the phase switch  and its amplitude $f$ are tuned such  that  it acts as $\pi$-pulse for a qubit of the median frequency $h$. It follows from  Rabi physics that rectangular  $\pi$-pulse is realized if  the qubit's  and carrying frequencies are brought into the resonance and if the following relation holds:
   \begin{equation}
   T=\frac{2\pi }{f}.
   \end{equation}

The further investigations  involve dynamics of  the Neel's  antiferromagnetic ordered state in a view of the Loschmidt echo. Also, the qubit-qubit correlations, spin-density wave decay and  fidelity, which is an overlap between initial and final states, are analyzed.

The quantum mechanical definition of the Loschmidt echo reads as
        \begin{equation}
     	S(t)= \Bigl|\langle \psi(0)| \mathcal{\bar T} e^{ {\rm i}\int\limits_0^t H(t') {\rm d}t' +\delta H t}\mathcal{T} e^{-{\rm i}\int\limits_0^t H(t') {\rm d}t'}|\psi(0) \rangle\Bigr| .
     \end{equation}
  We follow the notation of the Ref. \cite{Serbyn2017} where the initial state describes the spin-density wave 
\begin{equation}
   |\psi(0)\rangle=| \uparrow\downarrow\uparrow\downarrow . . . \downarrow\uparrow\rangle \label{neel}
   \end{equation}
and $\delta H = \Delta\epsilon \sigma^z_1$ is the perturbation term with $g=\Delta\epsilon$. The brackets $\langle \ \rangle$ also assumes the averaging over the disorder realizations. In order to analyze the entanglement we also calculate a correlation function of qubits polarization, similar to \cite{Pal2010}, in a translation invariant form
\begin{multline}
	C(d)=\sum\limits_{i}^{L}  \Bigl[ \langle \psi(t)| \sigma^z_i\sigma^z_{i+d}|\psi(t) \rangle - \\ - \langle \psi(t)| \sigma^z_i|\psi(t)\rangle\langle \psi(t)|\sigma^z_{i+d}|\psi(t) \rangle \Bigr],
\end{multline}
where the periodic boundary conditions for $i=1,L$ are applied.

We calculate the fidelity function $\alpha(t)$ and the spin-density-wave amplitude $A(t)$ defined as
\begin{equation}
	\alpha(t)=|\langle \psi(0)|\psi(t) \rangle|
\end{equation}
and  
\begin{equation}
	A(t)=\Bigl|\sum\limits_{i}^{L} (-1)^i \langle \psi(t)| \sigma^z_i|\psi(t) \rangle\Bigr|.
\end{equation}
 
\section{Results for Loschmidt echo and qubit-qubit correlations}
\label{results}
\begin{figure}[h]
	\includegraphics[width=1\linewidth]{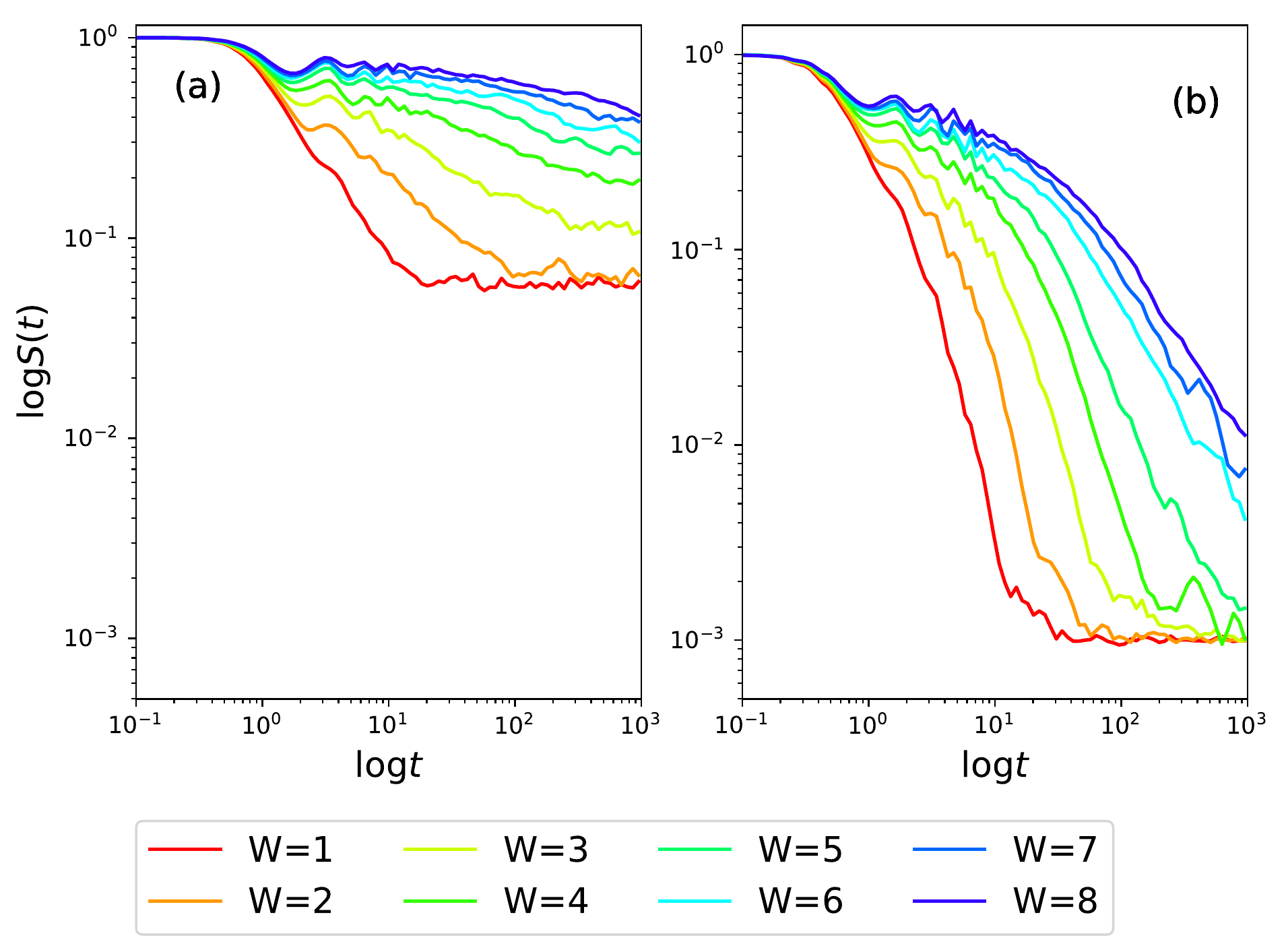}
	\caption{ 
		(a)  Results for the Loschmidt echo $S(t)$ for the Hamiltonian (\ref{hamiltonian-0})  without the driving, $f=0$, where the disorder strengths take  the values of  $W=1, \ ...\ 8$. The data is plotted in log-log scale for 8 qubits.   The MBL phase with the power-law   echo (linear sectors) in all of the curves is attributed to the 	initial Neel state which is a low lying excitation and does not show ergodicity even with low disorder. 
		(b) Results for   driven system,  $f=1$,   which demonstrate  many-body synchronization  and the suppression of the MBL phase. This latter is seen from the vanishing of the linear sectors of the echoes.   } \label{results-echo}
\end{figure}

 In this part we analyze numerically the effect of the transversal driving  on MBL phase by means   of the Loschmidt echo   $S(t)$ averaged by  disorder realizations.  
  In Fig. \ref{results-echo}(a) the echo is shown for the steady state Hamiltonian (\ref{hamiltonian-0}) without the driving.
  The  exchange constant in simulations is $J=1/2$ and the number of qubits is $L=8$. Increasing the disorder parameter $W$  from $W=1$ (lower red curve) to $W=8$ (upper magenta curve)  we observe  the decreasing of the decay rate (this data is in correspondence with the results of Ref. \cite{Serbyn2017}).  In the double logarithmic scale all of the curves in Fig. \ref{results-echo}(a) have linear sectors. This stands for  a power-law decay of the Loschmidt echo and indicates the   MBL phase for all of the $W$. There are  no curves with sub-power-law decay because the initial Neel state   $|\psi(t) \rangle$ introduced in (\ref{neel}) is a low lying state on energy scale which does not show ergodicity even with low disorder.
 Appearing of the driving, described by the Hamiltonian (\ref{hamiltonian-driving}), brings ergodicity.  The  data is shown in Fig. \ref{results-echo}(b) where the  driving strength   in simulations   is set to $f=1$. The dynamic transition into the ergodic phase  is seen by the disappearing of the linear sectors in the curves.   We associate this phenomena with the many-body synchronization of the  interacting   qubits. The value of the entropy saturation for long times in the  driven system, Fig. \ref{results-echo}(b), is two orders less than for $f=0$, Fig. \ref{results-echo}(a). This follows from   the increase of an effective temperature induced by  the driving.

\begin{figure}[h]
	\includegraphics[width=1\linewidth]{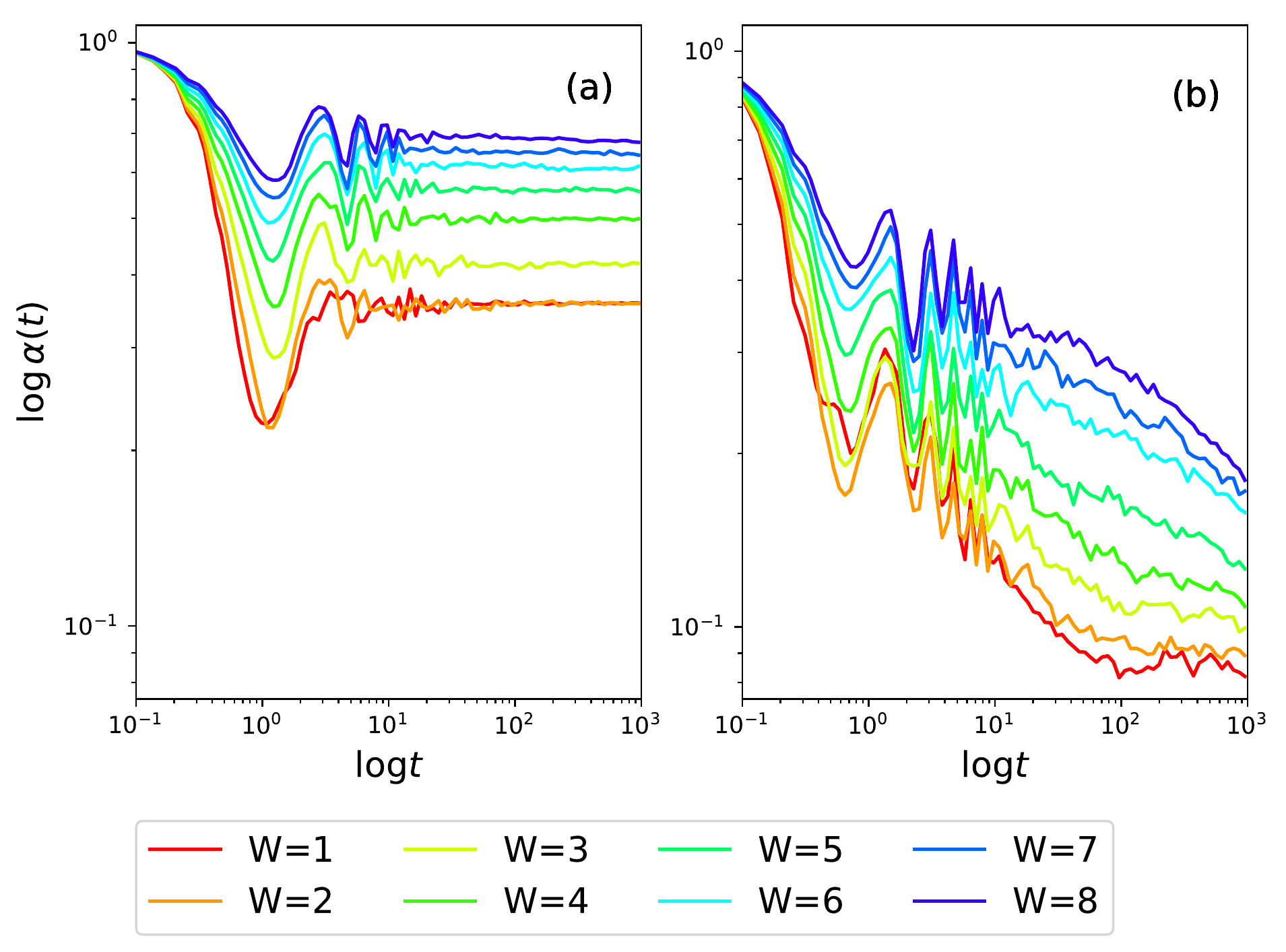}
	\caption{Results for the  fidelity $\alpha(t)$: (a) in steady state system without the driving, $f=0$, and  (b) in driven system, $f=1$. } \label{results-alpha}
\end{figure}

\begin{figure}[h]
	\includegraphics[width=1\linewidth]{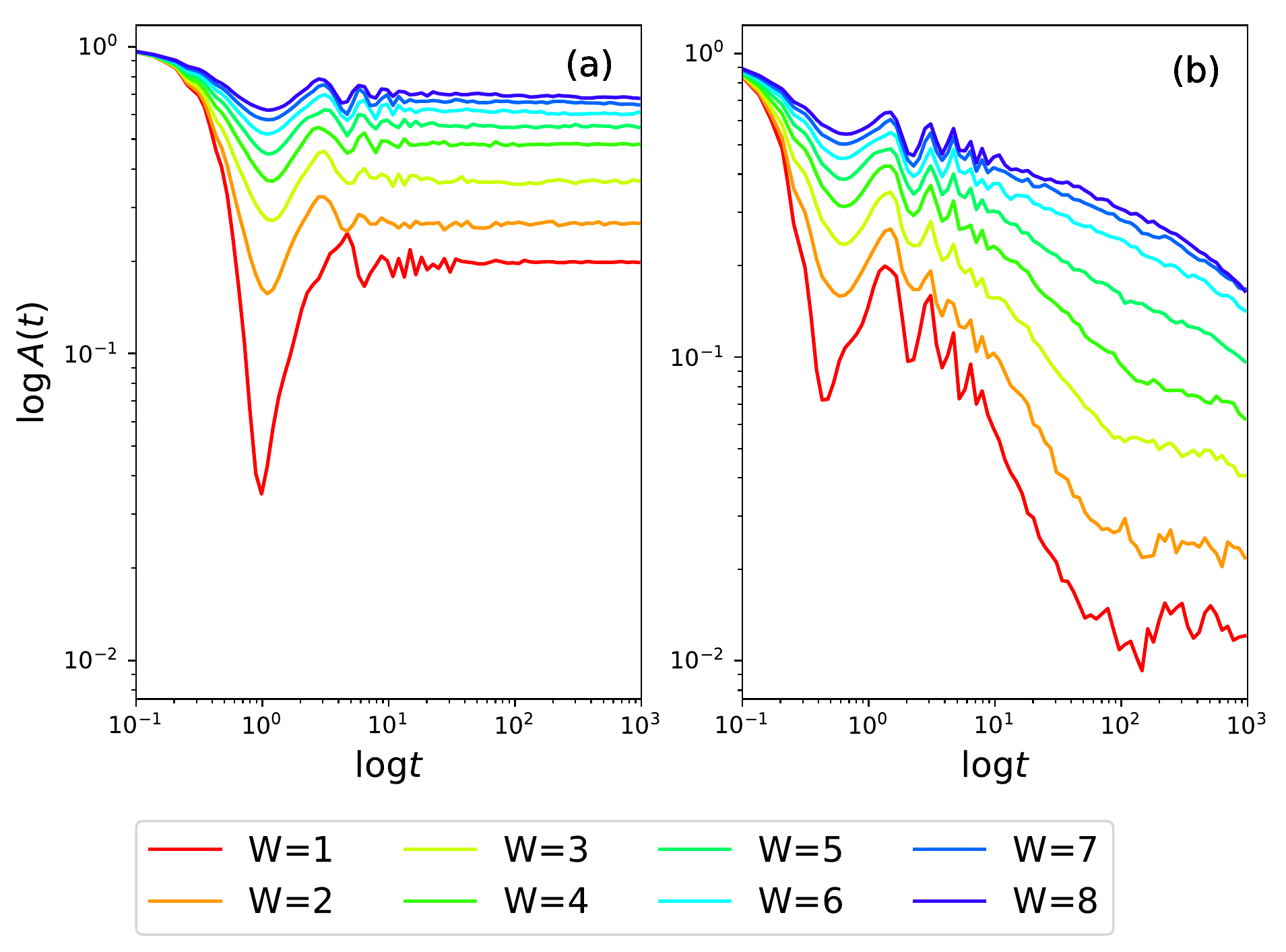}
	\caption{Results for the antiferromagnetic order parameter $A(t)$: (a) in steady state system without the driving, $f=0$, and  (b) in driven system, $f=1$. } \label{results-A}
\end{figure}

\begin{figure}[h]
	\includegraphics[width=1\linewidth]{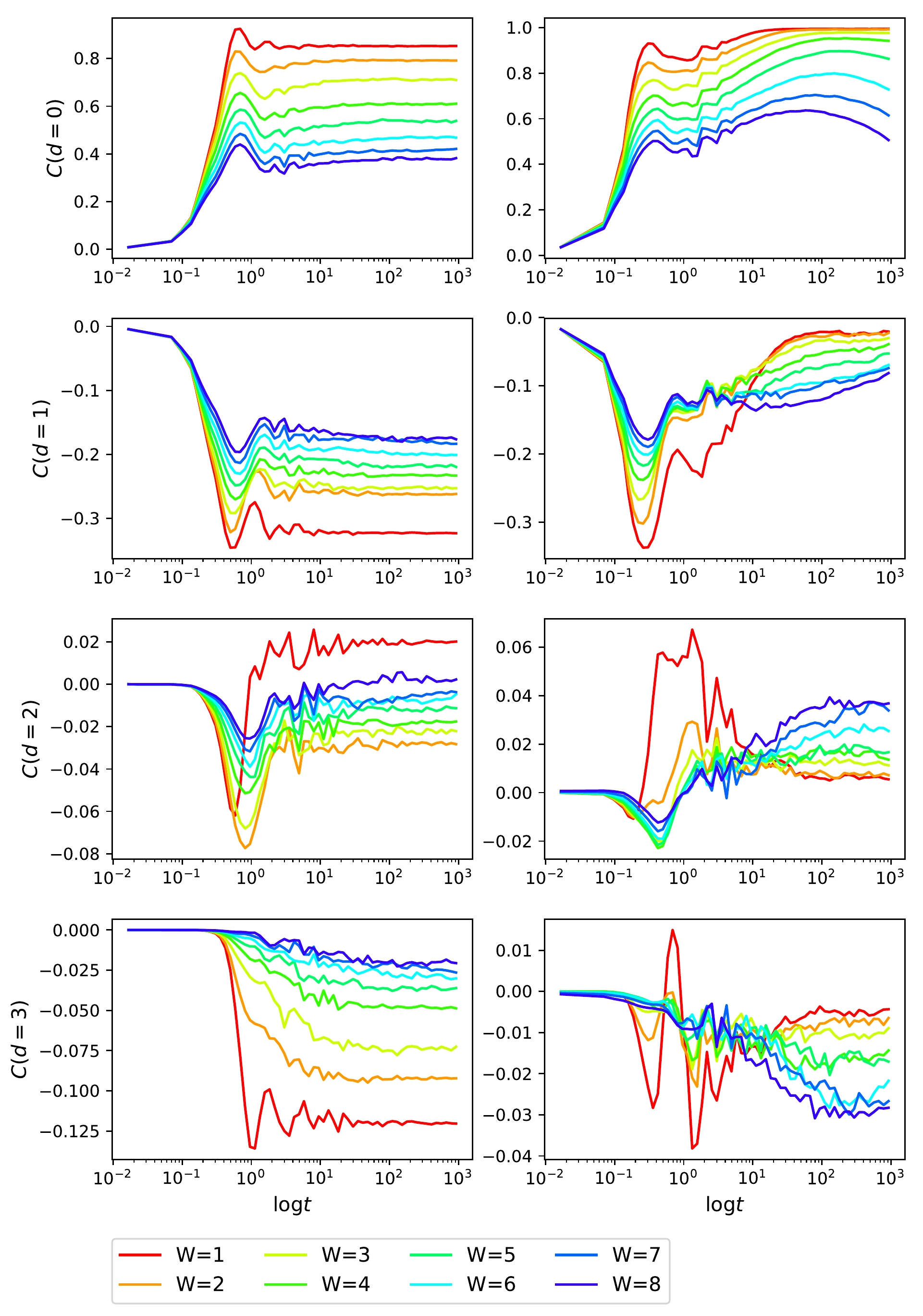}
	\caption{Results for the correlation function of qubits polarization $C(d)$ for different inter-qubit distances $d$: (left column) in steady state system without the driving, $f=0$, and  (right column) in driven system $f=1$.    } \label{results-C}
\end{figure}

The second part of the results is related to  the evolution  dynamics of the   Neel state in terms of the   fidelity and antiferromagnetic order parameter, $\alpha(t)$ and $A(t)$. As it is shown in Figs. \ref{results-alpha}(a) and \ref{results-A}(a),  if the driving is zero, $f=0$, then the Neel state evolution has a saturation to the finite values of $\alpha(t)$ and $A(t)$. The local momenta are formed in the system and do not decay in this case. Indeed, the switching on of the driving $f=1$, see Figs. \ref{results-alpha}(b) and \ref{results-A}(b),  reveals their decay and means that system is in ergodic phase. In contrast to 'single-particle'   characteristics  $\alpha (t)$ and $A(t)$, the correlator $C(d)$ shows more complicated behavior. It does not decay for the  driven system, as presented in the right column of Fig. \ref{results-C}.   
The correlator $C(d)$  has finite  value for low distances $d=1,2$  and zero driving, see Figs.  \ref{results-C} in left column. In the  non-zero driving regime, Figs.  \ref{results-C} in right column, all correlations are actually absent for long times.

\section{Discussion}
\label{discussion}

Being motivated by the work  \cite{Lukin},  where the  dynamical transition from the ergodic state into the MBL phase has been predicted for  qubits with switchable interaction parameters,  we searched for a dual regime with a controllable transition out of the  MBL phase.
Implementation of this transition is relevant for state-of-the-art   quantum interfaces where a large amount of controllable qubits  are integrated with each other. If a dynamical transition is possible then it would provide a method of   the disorder suppression. The unavoidable disorder in qubit frequencies, along with the decoherence, is always present   and  limits   their  collective performance.
  In this Letter we proposed the scenario of a controllable transition out of the MBL phase. It is realized as the driving
 of a transversal type (as Zeeman field in $xy$-plane in  terms of spin models), which has the periodic switching of its phase. Namely, we show that the applying of $\pi$-pulses with altering phases eliminates the signatures of the the MBL phase  and, consequently, can be considered as the {\it many-body synchronization} effect.
 Our numerical studies involve the  Loschmidt echo technique, which offers a powerful tool to probe the MBL phase, and the analysis of correlation functions and decay dynamics of the spin-density-wave excitations in the qubit ensemble.  
 It should be emphasized, that being applied to all the qubits simultaneously, the sequence of $\pi$-pulses    disregards a microscopic details of the disorder in qubits  frequencies.    We expect  that this transition offers a technique for an effective suppression of  inhomogeneous broadening in the ensemble.
 
 \section{Acknowledgments}
 The research was funded by the Russian Science Foundation under Grant No. 16-12-00095.


\begin{thebibliography}{49}
\expandafter\ifx\csname natexlab\endcsname\relax\def\natexlab#1{#1}\fi
\expandafter\ifx\csname bibnamefont\endcsname\relax
  \def\bibnamefont#1{#1}\fi
\expandafter\ifx\csname bibfnamefont\endcsname\relax
  \def\bibfnamefont#1{#1}\fi
\expandafter\ifx\csname citenamefont\endcsname\relax
  \def\citenamefont#1{#1}\fi
\expandafter\ifx\csname url\endcsname\relax
  \def\url#1{\texttt{#1}}\fi
\expandafter\ifx\csname urlprefix\endcsname\relax\def\urlprefix{URL }\fi
\providecommand{\bibinfo}[2]{#2}
\providecommand{\eprint}[2][]{\url{#2}}

\bibitem[{\citenamefont{Srednicki}(1994)}]{Srednicki}
\bibinfo{author}{\bibfnamefont{M.}~\bibnamefont{Srednicki}},
  \bibinfo{journal}{Phys. Rev. E} \textbf{\bibinfo{volume}{50}},
  \bibinfo{pages}{888} (\bibinfo{year}{1994}),
  \urlprefix\url{https://link.aps.org/doi/10.1103/PhysRevE.50.888}.

\bibitem[{\citenamefont{Deutsch}(1991)}]{Deutsch}
\bibinfo{author}{\bibfnamefont{J.~M.} \bibnamefont{Deutsch}},
  \bibinfo{journal}{Phys. Rev. A} \textbf{\bibinfo{volume}{43}},
  \bibinfo{pages}{2046} (\bibinfo{year}{1991}),
  \urlprefix\url{https://link.aps.org/doi/10.1103/PhysRevA.43.2046}.

\bibitem[{\citenamefont{Pal and Huse}(2010)}]{Pal2010}
\bibinfo{author}{\bibfnamefont{A.}~\bibnamefont{Pal}} \bibnamefont{and}
  \bibinfo{author}{\bibfnamefont{D.~A.} \bibnamefont{Huse}},
  \bibinfo{journal}{Phys. Rev. B} \textbf{\bibinfo{volume}{82}},
  \bibinfo{pages}{174411} (\bibinfo{year}{2010}),
  \urlprefix\url{https://link.aps.org/doi/10.1103/PhysRevB.82.174411}.

\bibitem[{\citenamefont{Gornyi et~al.}(2005)\citenamefont{Gornyi, Mirlin, and
  Polyakov}}]{Gornyi2005}
\bibinfo{author}{\bibfnamefont{I.~V.} \bibnamefont{Gornyi}},
  \bibinfo{author}{\bibfnamefont{A.~D.} \bibnamefont{Mirlin}},
  \bibnamefont{and} \bibinfo{author}{\bibfnamefont{D.~G.}
  \bibnamefont{Polyakov}}, \bibinfo{journal}{Phys. Rev. Lett.}
  \textbf{\bibinfo{volume}{95}}, \bibinfo{pages}{206603}
  (\bibinfo{year}{2005}),
  \urlprefix\url{https://link.aps.org/doi/10.1103/PhysRevLett.95.206603}.

\bibitem[{\citenamefont{Basko et~al.}(2006)\citenamefont{Basko, Aleiner, and
  Altshuler}}]{Basko2006}
\bibinfo{author}{\bibfnamefont{D.}~\bibnamefont{Basko}},
  \bibinfo{author}{\bibfnamefont{I.}~\bibnamefont{Aleiner}}, \bibnamefont{and}
  \bibinfo{author}{\bibfnamefont{B.}~\bibnamefont{Altshuler}},
  \bibinfo{journal}{Annals of Physics} \textbf{\bibinfo{volume}{321}},
  \bibinfo{pages}{1126 } (\bibinfo{year}{2006}), ISSN
  \bibinfo{issn}{0003-4916},
  \urlprefix\url{http://www.sciencedirect.com/science/article/pii/S0003491605002630}.

\bibitem[{\citenamefont{Oganesyan and Huse}(2007)}]{Oganesyan2007}
\bibinfo{author}{\bibfnamefont{V.}~\bibnamefont{Oganesyan}} \bibnamefont{and}
  \bibinfo{author}{\bibfnamefont{D.~A.} \bibnamefont{Huse}},
  \bibinfo{journal}{Phys. Rev. B} \textbf{\bibinfo{volume}{75}},
  \bibinfo{pages}{155111} (\bibinfo{year}{2007}),
  \urlprefix\url{https://link.aps.org/doi/10.1103/PhysRevB.75.155111}.

\bibitem[{\citenamefont{Nandkishore and Huse}(2015)}]{Nandkishore2015}
\bibinfo{author}{\bibfnamefont{R.}~\bibnamefont{Nandkishore}} \bibnamefont{and}
  \bibinfo{author}{\bibfnamefont{D.~A.} \bibnamefont{Huse}},
  \bibinfo{journal}{Annu. Rev. Condens. Matter Phys.}
  \textbf{\bibinfo{volume}{6}}, \bibinfo{pages}{15} (\bibinfo{year}{2015}).

\bibitem[{\citenamefont{Bennett et~al.}(1996)\citenamefont{Bennett, Bernstein,
  Popescu, and Schumacher}}]{Bennett1996}
\bibinfo{author}{\bibfnamefont{C.~H.} \bibnamefont{Bennett}},
  \bibinfo{author}{\bibfnamefont{H.~J.} \bibnamefont{Bernstein}},
  \bibinfo{author}{\bibfnamefont{S.}~\bibnamefont{Popescu}}, \bibnamefont{and}
  \bibinfo{author}{\bibfnamefont{B.}~\bibnamefont{Schumacher}},
  \bibinfo{journal}{Phys. Rev. A} \textbf{\bibinfo{volume}{53}},
  \bibinfo{pages}{2046} (\bibinfo{year}{1996}),
  \urlprefix\url{https://link.aps.org/doi/10.1103/PhysRevA.53.2046}.

\bibitem[{\citenamefont{Vidal et~al.}(2003)\citenamefont{Vidal, Latorre, Rico,
  and Kitaev}}]{Vidal2003}
\bibinfo{author}{\bibfnamefont{G.}~\bibnamefont{Vidal}},
  \bibinfo{author}{\bibfnamefont{J.~I.} \bibnamefont{Latorre}},
  \bibinfo{author}{\bibfnamefont{E.}~\bibnamefont{Rico}}, \bibnamefont{and}
  \bibinfo{author}{\bibfnamefont{A.}~\bibnamefont{Kitaev}},
  \bibinfo{journal}{Phys. Rev. Lett.} \textbf{\bibinfo{volume}{90}},
  \bibinfo{pages}{227902} (\bibinfo{year}{2003}),
  \urlprefix\url{https://link.aps.org/doi/10.1103/PhysRevLett.90.227902}.

\bibitem[{\citenamefont{Berkovits}(2012)}]{Berkovits2012}
\bibinfo{author}{\bibfnamefont{R.}~\bibnamefont{Berkovits}},
  \bibinfo{journal}{Phys. Rev. Lett.} \textbf{\bibinfo{volume}{108}},
  \bibinfo{pages}{176803} (\bibinfo{year}{2012}),
  \urlprefix\url{https://link.aps.org/doi/10.1103/PhysRevLett.108.176803}.

\bibitem[{\citenamefont{Bauer and Nayak}(2013)}]{Bauer2013}
\bibinfo{author}{\bibfnamefont{B.}~\bibnamefont{Bauer}} \bibnamefont{and}
  \bibinfo{author}{\bibfnamefont{C.}~\bibnamefont{Nayak}},
  \bibinfo{journal}{Journal of Statistical Mechanics: Theory and Experiment}
  \textbf{\bibinfo{volume}{2013}}, \bibinfo{pages}{P09005}
  (\bibinfo{year}{2013}),
  \urlprefix\url{http://stacks.iop.org/1742-5468/2013/i=09/a=P09005}.

\bibitem[{\citenamefont{Burmistrov et~al.}(2017)\citenamefont{Burmistrov,
  Tikhonov, Gornyi, and Mirlin}}]{BURMISTROV2017140}
\bibinfo{author}{\bibfnamefont{I.}~\bibnamefont{Burmistrov}},
  \bibinfo{author}{\bibfnamefont{K.}~\bibnamefont{Tikhonov}},
  \bibinfo{author}{\bibfnamefont{I.}~\bibnamefont{Gornyi}}, \bibnamefont{and}
  \bibinfo{author}{\bibfnamefont{A.}~\bibnamefont{Mirlin}},
  \bibinfo{journal}{Annals of Physics} \textbf{\bibinfo{volume}{383}},
  \bibinfo{pages}{140 } (\bibinfo{year}{2017}), ISSN \bibinfo{issn}{0003-4916},
  \urlprefix\url{http://www.sciencedirect.com/science/article/pii/S0003491617301392}.

\bibitem[{\citenamefont{Schreiber et~al.}(2015)\citenamefont{Schreiber,
  Hodgman, Bordia, L{\"u}schen, Fischer, Vosk, Altman, Schneider, and
  Bloch}}]{Schreiber842}
\bibinfo{author}{\bibfnamefont{M.}~\bibnamefont{Schreiber}},
  \bibinfo{author}{\bibfnamefont{S.~S.} \bibnamefont{Hodgman}},
  \bibinfo{author}{\bibfnamefont{P.}~\bibnamefont{Bordia}},
  \bibinfo{author}{\bibfnamefont{H.~P.} \bibnamefont{L{\"u}schen}},
  \bibinfo{author}{\bibfnamefont{M.~H.} \bibnamefont{Fischer}},
  \bibinfo{author}{\bibfnamefont{R.}~\bibnamefont{Vosk}},
  \bibinfo{author}{\bibfnamefont{E.}~\bibnamefont{Altman}},
  \bibinfo{author}{\bibfnamefont{U.}~\bibnamefont{Schneider}},
  \bibnamefont{and} \bibinfo{author}{\bibfnamefont{I.}~\bibnamefont{Bloch}},
  \bibinfo{journal}{Science} \textbf{\bibinfo{volume}{349}},
  \bibinfo{pages}{842} (\bibinfo{year}{2015}), ISSN \bibinfo{issn}{0036-8075},
  \eprint{http://science.sciencemag.org/content/349/6250/842.full.pdf},
  \urlprefix\url{http://science.sciencemag.org/content/349/6250/842}.

\bibitem[{\citenamefont{Choi et~al.}(2017{\natexlab{a}})\citenamefont{Choi,
  Choi, Landig, Kucsko, Zhou, Isoya, Jelezko, Onoda, Sumiya, Khemani
  et~al.}}]{Choi2017}
\bibinfo{author}{\bibfnamefont{S.}~\bibnamefont{Choi}},
  \bibinfo{author}{\bibfnamefont{J.}~\bibnamefont{Choi}},
  \bibinfo{author}{\bibfnamefont{R.}~\bibnamefont{Landig}},
  \bibinfo{author}{\bibfnamefont{G.}~\bibnamefont{Kucsko}},
  \bibinfo{author}{\bibfnamefont{H.}~\bibnamefont{Zhou}},
  \bibinfo{author}{\bibfnamefont{J.}~\bibnamefont{Isoya}},
  \bibinfo{author}{\bibfnamefont{F.}~\bibnamefont{Jelezko}},
  \bibinfo{author}{\bibfnamefont{S.}~\bibnamefont{Onoda}},
  \bibinfo{author}{\bibfnamefont{H.}~\bibnamefont{Sumiya}},
  \bibinfo{author}{\bibfnamefont{V.}~\bibnamefont{Khemani}},
  \bibnamefont{et~al.}, \bibinfo{journal}{Nature}
  \textbf{\bibinfo{volume}{543}}, \bibinfo{pages}{221}
  (\bibinfo{year}{2017}{\natexlab{a}}).

\bibitem[{\citenamefont{Bordia et~al.}(2017)\citenamefont{Bordia, L{\"u}schen,
  Schneider, Knap, and Bloch}}]{Bordia2017}
\bibinfo{author}{\bibfnamefont{P.}~\bibnamefont{Bordia}},
  \bibinfo{author}{\bibfnamefont{H.}~\bibnamefont{L{\"u}schen}},
  \bibinfo{author}{\bibfnamefont{U.}~\bibnamefont{Schneider}},
  \bibinfo{author}{\bibfnamefont{M.}~\bibnamefont{Knap}}, \bibnamefont{and}
  \bibinfo{author}{\bibfnamefont{I.}~\bibnamefont{Bloch}},
  \bibinfo{journal}{Nature Physics} \textbf{\bibinfo{volume}{13}},
  \bibinfo{pages}{460} (\bibinfo{year}{2017}).

\bibitem[{\citenamefont{Bernien et~al.}(2017)\citenamefont{Bernien, Schwartz,
  Keesling, Levine, Omran, Pichler, Choi, Zibrov, Endres, Greiner
  et~al.}}]{Bernien2017}
\bibinfo{author}{\bibfnamefont{H.}~\bibnamefont{Bernien}},
  \bibinfo{author}{\bibfnamefont{S.}~\bibnamefont{Schwartz}},
  \bibinfo{author}{\bibfnamefont{A.}~\bibnamefont{Keesling}},
  \bibinfo{author}{\bibfnamefont{H.}~\bibnamefont{Levine}},
  \bibinfo{author}{\bibfnamefont{A.}~\bibnamefont{Omran}},
  \bibinfo{author}{\bibfnamefont{H.}~\bibnamefont{Pichler}},
  \bibinfo{author}{\bibfnamefont{S.}~\bibnamefont{Choi}},
  \bibinfo{author}{\bibfnamefont{A.~S.} \bibnamefont{Zibrov}},
  \bibinfo{author}{\bibfnamefont{M.}~\bibnamefont{Endres}},
  \bibinfo{author}{\bibfnamefont{M.}~\bibnamefont{Greiner}},
  \bibnamefont{et~al.}, \bibinfo{journal}{arXiv preprint arXiv:1707.04344}
  (\bibinfo{year}{2017}).

\bibitem[{\citenamefont{\ifmmode \check{Z}\else
  \v{Z}\fi{}nidari\ifmmode~\check{c}\else \v{c}\fi{}
  et~al.}(2008)\citenamefont{\ifmmode \check{Z}\else
  \v{Z}\fi{}nidari\ifmmode~\check{c}\else \v{c}\fi{}, Prosen, and
  Prelov\ifmmode~\check{s}\else \v{s}\fi{}ek}}]{Znidaric2008}
\bibinfo{author}{\bibfnamefont{M.}~\bibnamefont{\ifmmode \check{Z}\else
  \v{Z}\fi{}nidari\ifmmode~\check{c}\else \v{c}\fi{}}},
  \bibinfo{author}{\bibfnamefont{T.~c.~v.} \bibnamefont{Prosen}},
  \bibnamefont{and}
  \bibinfo{author}{\bibfnamefont{P.}~\bibnamefont{Prelov\ifmmode~\check{s}\else
  \v{s}\fi{}ek}}, \bibinfo{journal}{Phys. Rev. B}
  \textbf{\bibinfo{volume}{77}}, \bibinfo{pages}{064426}
  (\bibinfo{year}{2008}),
  \urlprefix\url{https://link.aps.org/doi/10.1103/PhysRevB.77.064426}.

\bibitem[{\citenamefont{Bardarson et~al.}(2012)\citenamefont{Bardarson,
  Pollmann, and Moore}}]{Bardarson2012}
\bibinfo{author}{\bibfnamefont{J.~H.} \bibnamefont{Bardarson}},
  \bibinfo{author}{\bibfnamefont{F.}~\bibnamefont{Pollmann}}, \bibnamefont{and}
  \bibinfo{author}{\bibfnamefont{J.~E.} \bibnamefont{Moore}},
  \bibinfo{journal}{Phys. Rev. Lett.} \textbf{\bibinfo{volume}{109}},
  \bibinfo{pages}{017202} (\bibinfo{year}{2012}),
  \urlprefix\url{https://link.aps.org/doi/10.1103/PhysRevLett.109.017202}.

\bibitem[{\citenamefont{Huse et~al.}(2014)\citenamefont{Huse, Nandkishore, and
  Oganesyan}}]{Huse2014}
\bibinfo{author}{\bibfnamefont{D.~A.} \bibnamefont{Huse}},
  \bibinfo{author}{\bibfnamefont{R.}~\bibnamefont{Nandkishore}},
  \bibnamefont{and}
  \bibinfo{author}{\bibfnamefont{V.}~\bibnamefont{Oganesyan}},
  \bibinfo{journal}{Phys. Rev. B} \textbf{\bibinfo{volume}{90}},
  \bibinfo{pages}{174202} (\bibinfo{year}{2014}),
  \urlprefix\url{https://link.aps.org/doi/10.1103/PhysRevB.90.174202}.

\bibitem[{\citenamefont{Vasseur et~al.}(2015)\citenamefont{Vasseur,
  Parameswaran, and Moore}}]{Vasseur2015}
\bibinfo{author}{\bibfnamefont{R.}~\bibnamefont{Vasseur}},
  \bibinfo{author}{\bibfnamefont{S.~A.} \bibnamefont{Parameswaran}},
  \bibnamefont{and} \bibinfo{author}{\bibfnamefont{J.~E.} \bibnamefont{Moore}},
  \bibinfo{journal}{Phys. Rev. B} \textbf{\bibinfo{volume}{91}},
  \bibinfo{pages}{140202} (\bibinfo{year}{2015}),
  \urlprefix\url{https://link.aps.org/doi/10.1103/PhysRevB.91.140202}.

\bibitem[{\citenamefont{Serbyn et~al.}(2013)\citenamefont{Serbyn,
  Papi\ifmmode~\acute{c}\else \'{c}\fi{}, and Abanin}}]{Serbyn2013}
\bibinfo{author}{\bibfnamefont{M.}~\bibnamefont{Serbyn}},
  \bibinfo{author}{\bibfnamefont{Z.}~\bibnamefont{Papi\ifmmode~\acute{c}\else
  \'{c}\fi{}}}, \bibnamefont{and} \bibinfo{author}{\bibfnamefont{D.~A.}
  \bibnamefont{Abanin}}, \bibinfo{journal}{Phys. Rev. Lett.}
  \textbf{\bibinfo{volume}{111}}, \bibinfo{pages}{127201}
  (\bibinfo{year}{2013}),
  \urlprefix\url{https://link.aps.org/doi/10.1103/PhysRevLett.111.127201}.

\bibitem[{\citenamefont{Serbyn et~al.}(2015)\citenamefont{Serbyn,
  Papi\ifmmode~\acute{c}\else \'{c}\fi{}, and Abanin}}]{Serbyn2015}
\bibinfo{author}{\bibfnamefont{M.}~\bibnamefont{Serbyn}},
  \bibinfo{author}{\bibfnamefont{Z.}~\bibnamefont{Papi\ifmmode~\acute{c}\else
  \'{c}\fi{}}}, \bibnamefont{and} \bibinfo{author}{\bibfnamefont{D.~A.}
  \bibnamefont{Abanin}}, \bibinfo{journal}{Phys. Rev. X}
  \textbf{\bibinfo{volume}{5}}, \bibinfo{pages}{041047} (\bibinfo{year}{2015}),
  \urlprefix\url{https://link.aps.org/doi/10.1103/PhysRevX.5.041047}.

\bibitem[{\citenamefont{Latorre and Riera}(2009)}]{Latorre2009}
\bibinfo{author}{\bibfnamefont{J.~I.} \bibnamefont{Latorre}} \bibnamefont{and}
  \bibinfo{author}{\bibfnamefont{A.}~\bibnamefont{Riera}},
  \bibinfo{journal}{Journal of Physics A: Mathematical and Theoretical}
  \textbf{\bibinfo{volume}{42}}, \bibinfo{pages}{504002}
  (\bibinfo{year}{2009}),
  \urlprefix\url{http://stacks.iop.org/1751-8121/42/i=50/a=504002}.

\bibitem[{\citenamefont{Vitagliano et~al.}(2010)\citenamefont{Vitagliano,
  Riera, and Latorre}}]{Vitagliano2010}
\bibinfo{author}{\bibfnamefont{G.}~\bibnamefont{Vitagliano}},
  \bibinfo{author}{\bibfnamefont{A.}~\bibnamefont{Riera}}, \bibnamefont{and}
  \bibinfo{author}{\bibfnamefont{J.~I.} \bibnamefont{Latorre}},
  \bibinfo{journal}{New Journal of Physics} \textbf{\bibinfo{volume}{12}},
  \bibinfo{pages}{113049} (\bibinfo{year}{2010}),
  \urlprefix\url{http://stacks.iop.org/1367-2630/12/i=11/a=113049}.

\bibitem[{\citenamefont{Goussev et~al.}(2012)\citenamefont{Goussev, Jalabert,
  Pastawski, and Wisniacki}}]{Goussev:2012}
\bibinfo{author}{\bibfnamefont{A.}~\bibnamefont{Goussev}},
  \bibinfo{author}{\bibfnamefont{R.~A.} \bibnamefont{Jalabert}},
  \bibinfo{author}{\bibfnamefont{H.~M.} \bibnamefont{Pastawski}},
  \bibnamefont{and} \bibinfo{author}{\bibfnamefont{D.~A.}
  \bibnamefont{Wisniacki}}, \bibinfo{journal}{Scholarpedia}
  \textbf{\bibinfo{volume}{7}}, \bibinfo{pages}{11687} (\bibinfo{year}{2012}),
  \bibinfo{note}{revision \#127578}.

\bibitem[{\citenamefont{Serbyn and Abanin}(2017)}]{Serbyn2017}
\bibinfo{author}{\bibfnamefont{M.}~\bibnamefont{Serbyn}} \bibnamefont{and}
  \bibinfo{author}{\bibfnamefont{D.~A.} \bibnamefont{Abanin}},
  \bibinfo{journal}{Phys. Rev. B} \textbf{\bibinfo{volume}{96}},
  \bibinfo{pages}{014202} (\bibinfo{year}{2017}),
  \urlprefix\url{https://link.aps.org/doi/10.1103/PhysRevB.96.014202}.

\bibitem[{\citenamefont{Beaudoin et~al.}(2012)\citenamefont{Beaudoin, da~Silva,
  Dutton, and Blais}}]{Beaudoin2012}
\bibinfo{author}{\bibfnamefont{F.}~\bibnamefont{Beaudoin}},
  \bibinfo{author}{\bibfnamefont{M.~P.} \bibnamefont{da~Silva}},
  \bibinfo{author}{\bibfnamefont{Z.}~\bibnamefont{Dutton}}, \bibnamefont{and}
  \bibinfo{author}{\bibfnamefont{A.}~\bibnamefont{Blais}},
  \bibinfo{journal}{Phys. Rev. A} \textbf{\bibinfo{volume}{86}},
  \bibinfo{pages}{022305} (\bibinfo{year}{2012}),
  \urlprefix\url{https://link.aps.org/doi/10.1103/PhysRevA.86.022305}.

\bibitem[{\citenamefont{Srinivasan et~al.}(2011)\citenamefont{Srinivasan,
  Hoffman, Gambetta, and Houck}}]{Srinivasan2011}
\bibinfo{author}{\bibfnamefont{S.}~\bibnamefont{Srinivasan}},
  \bibinfo{author}{\bibfnamefont{A.}~\bibnamefont{Hoffman}},
  \bibinfo{author}{\bibfnamefont{J.}~\bibnamefont{Gambetta}}, \bibnamefont{and}
  \bibinfo{author}{\bibfnamefont{A.}~\bibnamefont{Houck}},
  \bibinfo{journal}{Physical review letters} \textbf{\bibinfo{volume}{106}},
  \bibinfo{pages}{083601} (\bibinfo{year}{2011}).

\bibitem[{\citenamefont{Hoffman et~al.}(2011)\citenamefont{Hoffman, Srinivasan,
  Gambetta, and Houck}}]{Hoffman2011}
\bibinfo{author}{\bibfnamefont{A.~J.} \bibnamefont{Hoffman}},
  \bibinfo{author}{\bibfnamefont{S.~J.} \bibnamefont{Srinivasan}},
  \bibinfo{author}{\bibfnamefont{J.~M.} \bibnamefont{Gambetta}},
  \bibnamefont{and} \bibinfo{author}{\bibfnamefont{A.~A.} \bibnamefont{Houck}},
  \bibinfo{journal}{Physical Review B} \textbf{\bibinfo{volume}{84}},
  \bibinfo{pages}{184515} (\bibinfo{year}{2011}).

\bibitem[{\citenamefont{Chen et~al.}(2014)\citenamefont{Chen, Neill, Roushan,
  Leung, Fang, Barends, Kelly, Campbell, Chen, Chiaro et~al.}}]{Chen2014}
\bibinfo{author}{\bibfnamefont{Y.}~\bibnamefont{Chen}},
  \bibinfo{author}{\bibfnamefont{C.}~\bibnamefont{Neill}},
  \bibinfo{author}{\bibfnamefont{P.}~\bibnamefont{Roushan}},
  \bibinfo{author}{\bibfnamefont{N.}~\bibnamefont{Leung}},
  \bibinfo{author}{\bibfnamefont{M.}~\bibnamefont{Fang}},
  \bibinfo{author}{\bibfnamefont{R.}~\bibnamefont{Barends}},
  \bibinfo{author}{\bibfnamefont{J.}~\bibnamefont{Kelly}},
  \bibinfo{author}{\bibfnamefont{B.}~\bibnamefont{Campbell}},
  \bibinfo{author}{\bibfnamefont{Z.}~\bibnamefont{Chen}},
  \bibinfo{author}{\bibfnamefont{B.}~\bibnamefont{Chiaro}},
  \bibnamefont{et~al.}, \bibinfo{journal}{Physical review letters}
  \textbf{\bibinfo{volume}{113}}, \bibinfo{pages}{220502}
  (\bibinfo{year}{2014}).

\bibitem[{\citenamefont{Zeytino{\u{g}}lu
  et~al.}(2015)\citenamefont{Zeytino{\u{g}}lu, Pechal, Berger, Abdumalikov~Jr,
  Wallraff, and Filipp}}]{Zeytinouglu2015}
\bibinfo{author}{\bibfnamefont{S.}~\bibnamefont{Zeytino{\u{g}}lu}},
  \bibinfo{author}{\bibfnamefont{M.}~\bibnamefont{Pechal}},
  \bibinfo{author}{\bibfnamefont{S.}~\bibnamefont{Berger}},
  \bibinfo{author}{\bibfnamefont{A.}~\bibnamefont{Abdumalikov~Jr}},
  \bibinfo{author}{\bibfnamefont{A.}~\bibnamefont{Wallraff}}, \bibnamefont{and}
  \bibinfo{author}{\bibfnamefont{S.}~\bibnamefont{Filipp}},
  \bibinfo{journal}{Physical Review A} \textbf{\bibinfo{volume}{91}},
  \bibinfo{pages}{043846} (\bibinfo{year}{2015}).

\bibitem[{\citenamefont{Braum\"uller et~al.}(2015)\citenamefont{Braum\"uller,
  Cramer, Schl\"or, Rotzinger, Radtke, Lukashenko, Yang, Skacel, Probst,
  Marthaler et~al.}}]{Braumueller2015}
\bibinfo{author}{\bibfnamefont{J.}~\bibnamefont{Braum\"uller}},
  \bibinfo{author}{\bibfnamefont{J.}~\bibnamefont{Cramer}},
  \bibinfo{author}{\bibfnamefont{S.}~\bibnamefont{Schl\"or}},
  \bibinfo{author}{\bibfnamefont{H.}~\bibnamefont{Rotzinger}},
  \bibinfo{author}{\bibfnamefont{L.}~\bibnamefont{Radtke}},
  \bibinfo{author}{\bibfnamefont{A.}~\bibnamefont{Lukashenko}},
  \bibinfo{author}{\bibfnamefont{P.}~\bibnamefont{Yang}},
  \bibinfo{author}{\bibfnamefont{S.~T.} \bibnamefont{Skacel}},
  \bibinfo{author}{\bibfnamefont{S.}~\bibnamefont{Probst}},
  \bibinfo{author}{\bibfnamefont{M.}~\bibnamefont{Marthaler}},
  \bibnamefont{et~al.}, \bibinfo{journal}{Phys. Rev. B}
  \textbf{\bibinfo{volume}{91}}, \bibinfo{pages}{054523}
  (\bibinfo{year}{2015}),
  \urlprefix\url{https://link.aps.org/doi/10.1103/PhysRevB.91.054523}.

\bibitem[{\citenamefont{Remizov
  et~al.}(2017{\natexlab{a}})\citenamefont{Remizov, Zhukov, Shapiro, Pogosov,
  and Lozovik}}]{Remizov2017}
\bibinfo{author}{\bibfnamefont{S.}~\bibnamefont{Remizov}},
  \bibinfo{author}{\bibfnamefont{A.}~\bibnamefont{Zhukov}},
  \bibinfo{author}{\bibfnamefont{D.}~\bibnamefont{Shapiro}},
  \bibinfo{author}{\bibfnamefont{W.}~\bibnamefont{Pogosov}}, \bibnamefont{and}
  \bibinfo{author}{\bibfnamefont{Y.~E.} \bibnamefont{Lozovik}},
  \bibinfo{journal}{Physical Review A} \textbf{\bibinfo{volume}{96}},
  \bibinfo{pages}{043870} (\bibinfo{year}{2017}{\natexlab{a}}).

\bibitem[{\citenamefont{Shapiro et~al.}(2015)\citenamefont{Shapiro, Zhukov,
  Pogosov, and Lozovik}}]{Shapiro2015}
\bibinfo{author}{\bibfnamefont{D.~S.} \bibnamefont{Shapiro}},
  \bibinfo{author}{\bibfnamefont{A.~A.} \bibnamefont{Zhukov}},
  \bibinfo{author}{\bibfnamefont{W.~V.} \bibnamefont{Pogosov}},
  \bibnamefont{and} \bibinfo{author}{\bibfnamefont{Y.~E.}
  \bibnamefont{Lozovik}}, \bibinfo{journal}{Phys. Rev. A}
  \textbf{\bibinfo{volume}{91}}, \bibinfo{pages}{063814}
  (\bibinfo{year}{2015}),
  \urlprefix\url{https://link.aps.org/doi/10.1103/PhysRevA.91.063814}.

\bibitem[{\citenamefont{Macha et~al.}(2014)\citenamefont{Macha, Oelsner,
  Reiner, Marthaler, Andr\'e, Sch\"on, H\"ubner, Meyer, Il'ichev, and
  Ustinov}}]{Macha2014}
\bibinfo{author}{\bibfnamefont{P.}~\bibnamefont{Macha}},
  \bibinfo{author}{\bibfnamefont{G.}~\bibnamefont{Oelsner}},
  \bibinfo{author}{\bibfnamefont{J.-M.} \bibnamefont{Reiner}},
  \bibinfo{author}{\bibfnamefont{M.}~\bibnamefont{Marthaler}},
  \bibinfo{author}{\bibfnamefont{S.}~\bibnamefont{Andr\'e}},
  \bibinfo{author}{\bibfnamefont{G.}~\bibnamefont{Sch\"on}},
  \bibinfo{author}{\bibfnamefont{U.}~\bibnamefont{H\"ubner}},
  \bibinfo{author}{\bibfnamefont{H.-G.} \bibnamefont{Meyer}},
  \bibinfo{author}{\bibfnamefont{E.}~\bibnamefont{Il'ichev}}, \bibnamefont{and}
  \bibinfo{author}{\bibfnamefont{A.~V.} \bibnamefont{Ustinov}},
  \bibinfo{journal}{Nature Communications} \textbf{\bibinfo{volume}{5}},
  \bibinfo{pages}{5146} (\bibinfo{year}{2014}).

\bibitem[{\citenamefont{Shulga et~al.}(2017)\citenamefont{Shulga, Yang,
  Fedorov, Fistul, Weides, and Ustinov}}]{Shulga2017}
\bibinfo{author}{\bibfnamefont{K.~V.} \bibnamefont{Shulga}},
  \bibinfo{author}{\bibfnamefont{P.}~\bibnamefont{Yang}},
  \bibinfo{author}{\bibfnamefont{G.~P.} \bibnamefont{Fedorov}},
  \bibinfo{author}{\bibfnamefont{M.~V.} \bibnamefont{Fistul}},
  \bibinfo{author}{\bibfnamefont{M.}~\bibnamefont{Weides}}, \bibnamefont{and}
  \bibinfo{author}{\bibfnamefont{A.~V.} \bibnamefont{Ustinov}},
  \bibinfo{journal}{JETP Letters} \textbf{\bibinfo{volume}{105}},
  \bibinfo{pages}{47} (\bibinfo{year}{2017}), ISSN \bibinfo{issn}{1090-6487},
  \urlprefix\url{https://doi.org/10.1134/S0021364017010143}.

\bibitem[{\citenamefont{Fistul}(2017)}]{fistul2017quantum}
\bibinfo{author}{\bibfnamefont{M.}~\bibnamefont{Fistul}},
  \bibinfo{journal}{Scientific Reports} \textbf{\bibinfo{volume}{7}}
  (\bibinfo{year}{2017}).

\bibitem[{\citenamefont{Zhang et~al.}(2017)\citenamefont{Zhang, Huang,
  Gershenson, and Bell}}]{Zhang2017}
\bibinfo{author}{\bibfnamefont{W.}~\bibnamefont{Zhang}},
  \bibinfo{author}{\bibfnamefont{W.}~\bibnamefont{Huang}},
  \bibinfo{author}{\bibfnamefont{M.~E.} \bibnamefont{Gershenson}},
  \bibnamefont{and} \bibinfo{author}{\bibfnamefont{M.~T.} \bibnamefont{Bell}},
  \bibinfo{journal}{Phys. Rev. Applied} \textbf{\bibinfo{volume}{8}},
  \bibinfo{pages}{051001} (\bibinfo{year}{2017}),
  \urlprefix\url{https://link.aps.org/doi/10.1103/PhysRevApplied.8.051001}.

\bibitem[{\citenamefont{Dutt et~al.}(2007)\citenamefont{Dutt, Childress, Jiang,
  Togan, Maze, Jelezko, Zibrov, Hemmer, and Lukin}}]{Dutt2007}
\bibinfo{author}{\bibfnamefont{M.~G.} \bibnamefont{Dutt}},
  \bibinfo{author}{\bibfnamefont{L.}~\bibnamefont{Childress}},
  \bibinfo{author}{\bibfnamefont{L.}~\bibnamefont{Jiang}},
  \bibinfo{author}{\bibfnamefont{E.}~\bibnamefont{Togan}},
  \bibinfo{author}{\bibfnamefont{J.}~\bibnamefont{Maze}},
  \bibinfo{author}{\bibfnamefont{F.}~\bibnamefont{Jelezko}},
  \bibinfo{author}{\bibfnamefont{A.}~\bibnamefont{Zibrov}},
  \bibinfo{author}{\bibfnamefont{P.}~\bibnamefont{Hemmer}}, \bibnamefont{and}
  \bibinfo{author}{\bibfnamefont{M.}~\bibnamefont{Lukin}},
  \bibinfo{journal}{Science} \textbf{\bibinfo{volume}{316}},
  \bibinfo{pages}{1312} (\bibinfo{year}{2007}).

\bibitem[{\citenamefont{Sandner et~al.}(2012)\citenamefont{Sandner, Ritsch,
  Ams\"uss, Koller, N\"obauer, Putz, Schmiedmayer, and Majer}}]{Sandner2012}
\bibinfo{author}{\bibfnamefont{K.}~\bibnamefont{Sandner}},
  \bibinfo{author}{\bibfnamefont{H.}~\bibnamefont{Ritsch}},
  \bibinfo{author}{\bibfnamefont{R.}~\bibnamefont{Ams\"uss}},
  \bibinfo{author}{\bibfnamefont{C.}~\bibnamefont{Koller}},
  \bibinfo{author}{\bibfnamefont{T.}~\bibnamefont{N\"obauer}},
  \bibinfo{author}{\bibfnamefont{S.}~\bibnamefont{Putz}},
  \bibinfo{author}{\bibfnamefont{J.}~\bibnamefont{Schmiedmayer}},
  \bibnamefont{and} \bibinfo{author}{\bibfnamefont{J.}~\bibnamefont{Majer}},
  \bibinfo{journal}{Phys. Rev. A} \textbf{\bibinfo{volume}{85}},
  \bibinfo{pages}{053806} (\bibinfo{year}{2012}),
  \urlprefix\url{https://link.aps.org/doi/10.1103/PhysRevA.85.053806}.

\bibitem[{\citenamefont{Putz et~al.}(2014)\citenamefont{Putz, Krimer, Amsuess,
  Valookaran, Noebauer, Schmiedmayer, Rotter, and Majer}}]{Putz2014}
\bibinfo{author}{\bibfnamefont{S.}~\bibnamefont{Putz}},
  \bibinfo{author}{\bibfnamefont{D.~O.} \bibnamefont{Krimer}},
  \bibinfo{author}{\bibfnamefont{R.}~\bibnamefont{Amsuess}},
  \bibinfo{author}{\bibfnamefont{A.}~\bibnamefont{Valookaran}},
  \bibinfo{author}{\bibfnamefont{T.}~\bibnamefont{Noebauer}},
  \bibinfo{author}{\bibfnamefont{J.}~\bibnamefont{Schmiedmayer}},
  \bibinfo{author}{\bibfnamefont{S.}~\bibnamefont{Rotter}}, \bibnamefont{and}
  \bibinfo{author}{\bibfnamefont{J.}~\bibnamefont{Majer}},
  \bibinfo{journal}{Nature Physics} \textbf{\bibinfo{volume}{10}},
  \bibinfo{pages}{720} (\bibinfo{year}{2014}).

\bibitem[{\citenamefont{Remizov et~al.}(2015)\citenamefont{Remizov, Shapiro,
  and Rubtsov}}]{Remizov2015}
\bibinfo{author}{\bibfnamefont{S.~V.} \bibnamefont{Remizov}},
  \bibinfo{author}{\bibfnamefont{D.~S.} \bibnamefont{Shapiro}},
  \bibnamefont{and} \bibinfo{author}{\bibfnamefont{A.~N.}
  \bibnamefont{Rubtsov}}, \bibinfo{journal}{Phys. Rev. A}
  \textbf{\bibinfo{volume}{92}}, \bibinfo{pages}{053814}
  (\bibinfo{year}{2015}),
  \urlprefix\url{https://link.aps.org/doi/10.1103/PhysRevA.92.053814}.

\bibitem[{\citenamefont{Remizov
  et~al.}(2017{\natexlab{b}})\citenamefont{Remizov, Shapiro, and
  Rubtsov}}]{RemizovShapiroRubtsov2017}
\bibinfo{author}{\bibfnamefont{S.~V.} \bibnamefont{Remizov}},
  \bibinfo{author}{\bibfnamefont{D.~S.} \bibnamefont{Shapiro}},
  \bibnamefont{and} \bibinfo{author}{\bibfnamefont{A.~N.}
  \bibnamefont{Rubtsov}}, \bibinfo{journal}{JETP letters}
  \textbf{\bibinfo{volume}{105}}, \bibinfo{pages}{130}
  (\bibinfo{year}{2017}{\natexlab{b}}).

\bibitem[{\citenamefont{Choi et~al.}(2017{\natexlab{b}})\citenamefont{Choi,
  Abanin, and Lukin}}]{Lukin}
\bibinfo{author}{\bibfnamefont{S.}~\bibnamefont{Choi}},
  \bibinfo{author}{\bibfnamefont{D.~A.} \bibnamefont{Abanin}},
  \bibnamefont{and} \bibinfo{author}{\bibfnamefont{M.~D.} \bibnamefont{Lukin}},
  \bibinfo{journal}{arXiv:1703.03809}  (\bibinfo{year}{2017}{\natexlab{b}}).

\bibitem[{\citenamefont{Yamamoto et~al.}(2003)\citenamefont{Yamamoto, Pashkin,
  Astafiev, Nakamura, and Tsai}}]{Yamamoto2010}
\bibinfo{author}{\bibfnamefont{T.}~\bibnamefont{Yamamoto}},
  \bibinfo{author}{\bibfnamefont{Y.~A.} \bibnamefont{Pashkin}},
  \bibinfo{author}{\bibfnamefont{O.}~\bibnamefont{Astafiev}},
  \bibinfo{author}{\bibfnamefont{Y.}~\bibnamefont{Nakamura}}, \bibnamefont{and}
  \bibinfo{author}{\bibfnamefont{J.~S.} \bibnamefont{Tsai}},
  \bibinfo{journal}{Nature} \textbf{\bibinfo{volume}{425}},
  \bibinfo{pages}{941} (\bibinfo{year}{2003}),
  \urlprefix\url{http://dx.doi.org/10.1038/nature02015}.

\bibitem[{\citenamefont{Nation et~al.}(2012)\citenamefont{Nation, Johansson,
  Blencowe, and Nori}}]{nation2012colloquium}
\bibinfo{author}{\bibfnamefont{P.}~\bibnamefont{Nation}},
  \bibinfo{author}{\bibfnamefont{J.}~\bibnamefont{Johansson}},
  \bibinfo{author}{\bibfnamefont{M.}~\bibnamefont{Blencowe}}, \bibnamefont{and}
  \bibinfo{author}{\bibfnamefont{F.}~\bibnamefont{Nori}},
  \bibinfo{journal}{Reviews of Modern Physics} \textbf{\bibinfo{volume}{84}},
  \bibinfo{pages}{1} (\bibinfo{year}{2012}).

\bibitem[{\citenamefont{Blais et~al.}(2004)\citenamefont{Blais, Huang,
  Wallraff, Girvin, and Schoelkopf}}]{Blais2004}
\bibinfo{author}{\bibfnamefont{A.}~\bibnamefont{Blais}},
  \bibinfo{author}{\bibfnamefont{R.-S.} \bibnamefont{Huang}},
  \bibinfo{author}{\bibfnamefont{A.}~\bibnamefont{Wallraff}},
  \bibinfo{author}{\bibfnamefont{S.~M.} \bibnamefont{Girvin}},
  \bibnamefont{and} \bibinfo{author}{\bibfnamefont{R.~J.}
  \bibnamefont{Schoelkopf}}, \bibinfo{journal}{Phys. Rev. A}
  \textbf{\bibinfo{volume}{69}}, \bibinfo{pages}{062320}
  (\bibinfo{year}{2004}),
  \urlprefix\url{https://link.aps.org/doi/10.1103/PhysRevA.69.062320}.

\bibitem[{\citenamefont{Boissonneault et~al.}(2010)\citenamefont{Boissonneault,
  Gambetta, and Blais}}]{Boissonneault2010}
\bibinfo{author}{\bibfnamefont{M.}~\bibnamefont{Boissonneault}},
  \bibinfo{author}{\bibfnamefont{J.~M.} \bibnamefont{Gambetta}},
  \bibnamefont{and} \bibinfo{author}{\bibfnamefont{A.}~\bibnamefont{Blais}},
  \bibinfo{journal}{Phys. Rev. Lett.} \textbf{\bibinfo{volume}{105}},
  \bibinfo{pages}{100504} (\bibinfo{year}{2010}),
  \urlprefix\url{https://link.aps.org/doi/10.1103/PhysRevLett.105.100504}.

\bibitem[{\citenamefont{George et~al.}(2017)\citenamefont{George, Senior,
  Saira, Pekola, de~Graaf, Lindstr{\"o}m, and Pashkin}}]{George2017}
\bibinfo{author}{\bibfnamefont{R.~E.} \bibnamefont{George}},
  \bibinfo{author}{\bibfnamefont{J.}~\bibnamefont{Senior}},
  \bibinfo{author}{\bibfnamefont{O.-P.} \bibnamefont{Saira}},
  \bibinfo{author}{\bibfnamefont{J.~P.} \bibnamefont{Pekola}},
  \bibinfo{author}{\bibfnamefont{S.~E.} \bibnamefont{de~Graaf}},
  \bibinfo{author}{\bibfnamefont{T.}~\bibnamefont{Lindstr{\"o}m}},
  \bibnamefont{and} \bibinfo{author}{\bibfnamefont{Y.~A.}
  \bibnamefont{Pashkin}}, \bibinfo{journal}{Journal of Low Temperature Physics}
  \textbf{\bibinfo{volume}{189}}, \bibinfo{pages}{60} (\bibinfo{year}{2017}),
  ISSN \bibinfo{issn}{1573-7357},
  \urlprefix\url{https://doi.org/10.1007/s10909-017-1787-x}.

\end{thebibliography}
\end{document}